# Quantum diffusion regime of the charge transport in GdB$_6$ caused by electron and lattice instability.

Alexander P. Dudka [a,b], Olga N. Khrykina [a,b], Nadezhda B. Bolotina [a], Natalya Yu. Shitsevalova [c], Volodymyr B. Filipov [c], Mikhail A. Anisimov [b], Slavomir Gabani [d], Karol Flachbart [d], Nikolay E. Sluchanko [b,e,*]

[a] Shubnikov Institute of Crystallography of Federal Scientific Research Centre 'Crystallography and Photonics' of Russian Academy of Sciences, 59 Leninskiy Ave., 119333 Moscow, Russia

[b] Prokhorov General Physics Institute, Russian Academy of Sciences, 38 Vavilov Str., 119991 Moscow, Russia

[c] Frantsevich Institute for Problems of Materials Science, National Academy of Sciences of Ukraine, 3 Krzhyzhanovsky Str., 03142 Kiev, Ukraine

[d] Institute of Experimental Physics, Slovak Academy of Sciences, Košice, 04001 Slovak Republic

[e] Moscow Institute of Physics and Technology (State University), 9 Institutskiy Per., 141700 Dolgoprudny, Russia

*E-mail: nes@lt.gpi.ru

## Abstract

Based on accurate X-ray structure analysis of GdB$_6$ over the temperature range 85-300 K it has been shown that anomalously strong charge carrier scattering in the quantum diffusion regime of charge transport in this compound arises due to the formation of (*i*) dynamically coupled Gd$^{3+}$ dimers of about 3.3 Å in size with an energy of quasi-local oscillations ~ 7 - 8 meV and (*ii*) dynamic charge stripes along the [001] direction of the cubic lattice. It has been revealed that anharmonic approximation is useful when analyzing the static and dynamic components of the atomic displacement parameters of gadolinium. The barrier height of double-well potential of Gd$^{3+}$ ions was determined both from low-temperature heat capacity measurements and from the electron density distribution reconstructed from X-ray data.

PACS: 72.15.-v, 72.15.Eb, 63.20.Pw, 61.46.-w

2**1. Introduction**. A short time after the discovery of high-temperature superconductivity in cuprates, it became clear that the highest critical temperatures $T_c$ were observed in materials with a linear temperature dependence of resistivity $\rho(T)$ (see, for example, [1-5]), which could be extended to very high temperatures (~ 1000 K [1]) leading to violation of the Mott-Ioffe-Regel limit [6]. Later, a similar behavior of $\rho(T)$ was found in pnictides [7–8] and organic superconductors [9–10] as well as in many heavy fermionic metals and superconductors [11–12] including those located in the vicinity of quantum critical point (QCP) [13]. Many different mechanisms were proposed to explain this effect, including quantum critical theories [13] and more exotic approaches, but its nature is the subject of active debate so far. As shown by the analysis of experimental results obtained for $Sr_3Ru_2O_7$ in the vicinity of QCP [14], the linear dependence of resistivity can be characterized by the same carrier scattering frequency in conductors of different classes. A single description can thus be offered in terms of diffusion transport with a diffusion coefficient which has a value close to the quantum limit $D = \hbar/m^*$, where $\hbar$ is the Planck constant and $m^*$ is the effective mass of charge carriers [11].

Taking into account that all the above-listed strongly correlated electronic systems (SCES) have complicated electronic and crystal structures, it is of interest to study a single-crystal conductor with a cubic crystal structure and with $T$-linear resistivity. Fine structure details and mechanisms of carrier scattering, which cause the appearance of a diffusion regime of charge transport near the quantum limit, seem to be in this case the most promising for study. In this work, single crystals of gadolinium hexaboride $GdB_6$ with a simple cubic structure similar to that of $CaB_6$ (Fig.1a) and a T-linear resistivity $\rho(T)$ was opted to study fine details of the crystal and electronic structure, static and dynamic components of atomic displacement parameters and specific heat. Unlike some SCES located in the vicinity of QCP, $GdB_6$ is an antiferromagnetic (AF) metal (Néel temperature $T_N \approx 16$ K [15]), in which the majority of charge carriers (~70%) are non-equilibrium (hot) electrons, which participate in the formation of collective mode [16] already at room temperature. In the present study it is shown that dynamically coupled Gd dimers in combination with dynamic charge stripes are formed in $GdB_6$ with the temperature lowering (Fig.1b-d) and create very strong charge carrier scattering in the quantum diffusion regime.

**2. Experimental techniques**

High-quality single crystals of $GdB_6$ were grown by vertical crucible-free induction melting in an argon gas atmosphere. Preparation and characterization details of rare earth (RE) higher borides used in thermal and structural studies are described elsewhere [17-20]. Heat capacity was measured on a PPMS-9 (Quantum Design, Inc.) installation; measurements of resistance were performed in a four-terminal scheme with a direct current commutation. Single-crystal X-ray data were collected with extremely high accuracy at nine temperatures in the range 85 - 300 K using an Xcalibur diffractometer (Rigaku Oxford Diffraction) equipped with a CCD detector EOS S2.

## 3. Results and discussion

**3.1. Resistivity and specific heat.** Temperature dependences of resistivity $\rho(T)$ and specific heat $C(T)$ of GdB$_6$ are shown in Fig. 2a and 2b, correspondingly. As in [15], a linear $\rho(T)$ dependence is registered here in the range of 30-300 K and at $T_N \approx 16$ K strong anomalies are observed, which are associated with the AF transition. Similar data for a non-magnetic reference compound LaB$_6$ are presented, whose resistivity is 4–30 times lower than that of GdB$_6$. The comparison of low-temperature heat capacity curves (Fig. 2b) in the paramagnetic phase of GdB$_6$ and in diamagnetic LaB$_6$ leads to conclusion about significant differences that may be related, as Gd ions have no orbital momentum ($s = 7/2$, $l = 0$ for Gd$^{3+}$), to markedly different frequencies of quasi-local vibrations of La$^{3+}$ and Gd$^{3+}$ ions. The boron sub-lattice of these two RE hexaborides gives equal Debye contributions $C_D$ to heat capacity (their Debye temperature is $\theta_D = 1160$ K, Fig. 2c and Ref. 17), while the Einstein contributions $C_E$ from the RE ions are significantly different corresponding to Einstein temperatures $\theta_E$ (LaB$_6$) = 140 K [17, 21-22] and $\theta_E$ (GdB$_6$) = 91 K [23] (Fig. 2d). For GdB$_6$, additionally to a huge anomaly at $T_N \approx 16$ K we detect also a large-amplitude Schottky component $C_{Sch}$ (Fig. 2e) that corresponds to a barrier height of $\Delta E \approx 37$ K in double-well potential (DWP). It is worth noting that the Schottky contribution is about three orders of magnitude higher than it may be estimated from any possible magnetic anomaly caused by isolated Gd ions, therefore, it should be attributed to magneto-vibrating states of Gd-pairs (see below) located in the DWPs.

**3.2. Crystal structure.** The structure of GdB$_6$ was refined in cubic *Pm-3m* group at nine temperatures within 85 -300 K in the harmonic approximation of atomic vibrations. A unit cell of GdB$_6$ is presented in Fig. 1a. Central Gd atom is surrounded by eight B$_6$ octahedra situated at vertexes of the cubic lattice. Temperature dependent interatomic distances Gd–B, (B–B)$_{intra}$ in an octahedron and (B–B)$_{inter}$ between octahedra are presented in Fig. 3a.

The lattice parameters were additionally refined without symmetry-induced constraints. Temperature dependences of linear and angular lattice parameters are given in Figs. 3b and 3c, respectively (see designations in Fig. 1a). Differences in the lattice periods and angles, although very small, remain over the whole temperature range indicating violations of cubic symmetry, previously observed in the dodecaboride LuB$_{12}$ [24]. Since trigonal distortions of the LuB$_{12}$ lattice are highly likely to be the key factors determining the occurrence of dynamic charge stripes in higher borides, it is of interest to analyze difference Fourier maps of GdB$_6$ for possible symmetry violations as it was done in case of LuB$_{12}$. The Fourier synthesis of electron density (ED) does not require information on crystal symmetry being based on reflection intensities and atomic coordinates. This allows to analyze fine crystal structure details on symmetry-independent difference Fourier maps [18, 19, 24]. Visible (100), (010) and (001) faces of each cube on panels (a) - (c) of Fig. 4 contain Gd$^{3+}$ ions in vertexes of the unit cells. Each cube in Fig. 4 consists of eight unit cells for easier viewing of the filamentary structure of residual ED at temperatures 293, 200 and 90 K. Different values $\Delta g$ of residual ED within the layer 0.05(e Å$^{-3}$) < $\Delta g$ < 0.8$\Delta g_{max}$ are represented in Fig. 4 by different colors. This allows to visualize two characteristic properties: (1) already at room temperature, at distances of about ± 0.5 Å from the Gd site, and approximately in direction of the face diagonal of the unit cell, ED maxima predicting the formation of Gd–Gd dimers are observed; (2) with temperature decrease, the amplitude of the maxima increases significantly, and the observed maxima stretch into stripes, simultaneously turning approximately in the [0-13] direction in the (100) plane (Fig. 4c). As a result, zigzag shaped charge stripes are formed at low temperatures near the (100) plane, which are mainly oriented approximately along the $c$ axis (Fig. 4c). Residual ED distributions in the (100) plane of a single cell are shown in Fig. 1, panels (b) – (d) with an origin shifted by (½, ½, ½) as compared to Fig. 4. It should be emphasized that in the presence of cubic lattice distortions, the structure defects (boron vacancies [17, 22] and the structure disorder due to isotopic substitutions $^{10}$B–$^{11}$B in the boron sublattice (natural isotope distribution is 18.8% $^{10}$B and 81.2% $^{11}$B), the formation of selected direction along the charge stripes leads to the appearance of uniaxial anisotropy in these cubic crystals.

Evidently the appearance of additional ED at distances of about ± 0.5 Å from the Gd site should be analyzed in terms of double-well potentials and anharmonic vibrations of Gd-ions. One-particle potential $V^{OPP}(\mathbf{u})$ in a point distant by a vector $\mathbf{u}$ from a selected atom can be estimated from the formula [25]:

$$V^{OPP}(\mathbf{u}) = -k_B T\{\ln G(\mathbf{u}) - \ln G(\mathbf{u} = 0)\}, \quad (1)$$

where $G(\mathbf{u})$ is the generalized atomic probability density function. In our work, pseudo-potential curves were drawn along stripes in the (100) plane at each of nine temperatures using the formula (1) with symmetry-independent ED values of $g(\mathbf{u})$ instead of symmetry-restricted values of $G(\mathbf{u})$ in order to estimate the barrier heights $\Delta E(T)$ in the DWP of the Gd$^{3+}$ ion. Legality of such a replacement is discussed in [25]. Calculated from X-ray data the $\Delta E(T)$ dependence is presented in Fig. 2g where the DWP barrier height obtained from heat capacity measurements is also shown for comparison. The tendency to a decrease of Δ$E$(T) below 8 meV with the temperature lowering in the range 20-200K (Fig.2g) is in accordance with the softening of Gd quasilocal mode observed in [26, 27] by inelastic X-ray scattering. On the other hand, authors of [26-27] found that the L mode (the phonon branch propagating along the [110] axis) softened only by 9% from 300 K down to 20 K. For the longitudinal phonon mode along the [100] direction in GdB$_6$ a softening of about 13% was detected taking phonon energy value $E_{ph}$ ~ 5.7 meV at 20 K [28]. Besides, it was concluded in [28] that the phonon softening behavior was strongly anisotropic and indicating an anharmonic or shallow potential for Gd ions. It is worth noting that Raman scattering results [29-30] also indicate that the anharmonicity of the RE mode is anisotropic, that is, it depends on the vibrating and propagating directions of the RE mode in GdB$_6$. As follows from the difference ED maps, the most significant softening is developed in [0-13] direction in the (100) plane which corresponds to the configuration of the dynamic charge stripes in GdB$_6$ (Fig. 4c).

**3.3. Dynamic and static atomic displacements.** Atomic displacement parameters are components of the Debye–Waller factor $T$(**H**) that describes an atomic distribution near the lattice points being a part of the structure factor $F$(**H**) [31]. Harmonic vibrations are represented by the factor $T_{harm}$(**H**) = exp(-2π$^2U^{ij}a^ia^jh_ih_j$) where $U^{ij}$ is a second-rank tensor; $a^i$, $a^j$ and $h_i$, $h_j$, 1 ≤ $i, j$ ≤ 3, are respectively periods of reciprocal lattice and Miller indexes. An anharmonicity of atomic motion is accounted, if necessary, by an expansion of $T$(**H**) into the Gram − Charlier series with tensor coefficients of rank higher than two. The second-rank $U^{ij}$ tensor can be written in the form of symmetric matrix {$u_{ij}$} whose diagonal elements define an equivalent mean-square atomic displacement $u_{eq}$ = ($u_{11}$ + $u_{22}$ + $u_{33}$)/3 in any case, if even $T$(**H**) contains additional higher-rank tensors.

Difference Fourier maps shown above were built for the crystal structure of GdB$_6$ refined in the harmonic approximation of atomic vibrations (in fact, harmonic vibrations of Gd in special position 4$a$ of the *Fm-3m* group are isotropic), so that any asymmetry of the ED distribution could be well seen. Gd atoms loosely bound with boron cages and located in the large size cavities of a rigid covalent boron framework (the radius of B$_{24}$ cavity (~2 Å) exceeds strongly the Gd ionic radius r(Gd) ≈ 0.94 Å [23]) are usually considered as independent harmonic oscillators what allows to fit the temperature-dependent parameters $u_{eq}$(Gd) and $u_{eq}$(B) by Einstein (2) and Debye (3) formulas, respectively:

$$u_{eq}(\text{Gd}) = \frac{h^2}{4\pi^2 * k_B * m_a * \theta_E}\left(\frac{1}{2} + \frac{1}{exp\left(\frac{\theta_E}{T}-1\right)}\right) + \langle u^2 \rangle_{shift(Gd)} \qquad (2)$$

$$u_{eq}(\text{B}) = \frac{3*h^2}{4\pi^2 * k_B * m_a * \theta_D}\left(\frac{1}{4} + \left(\frac{T}{\theta_D}\right)^2 \int_0^{\frac{\theta_D}{T}} \frac{y}{\exp(y)-1} dy\right) + \langle u^2 \rangle_{shift(B)} \qquad (3)$$

where $h$, $k_B$ are Planck and Boltzmann constants, $m_a$ is atomic mass, $\theta_E$ and $\theta_D$ are the Einstein and Debye temperatures and $\langle u^2 \rangle_{shift}$ is the static mean-square displacement. In Eqs. (2) and (3) there are two dynamic components of the equivalent atomic displacement $u_{eq}$ which correspond to the thermal and zero-temperature vibrations of these atoms, and one static term that determines the temperature-independent shift of these atoms from their lattice sites. This static term arises mainly due to boron vacancies (the occurrence of about 1- 9% of vacancies at boron sites has been detected in $RB_6$ [22]) and substitutional disorder which is present in RE hexaborides with natural (18.8% $^{10}$B and 81.2% $^{11}$B) boron composition.

Difference Fourier maps of residual ED (Fig. 4) accurately indicate the appearance of Gd–Gd dimers at room temperature and this tendency increases strongly with the temperature lowering. The observed result is in accord with the dynamical Jahn-Teller (JT) model proposed by T. Kasuya for $GdB_6$ where each Gd atom distorts to a quasistable position along each [001] direction and the ground state is arranged as a linear combination of six equivalent sites keeping cubic symmetry (pair-distorted dynamic JT effect) [32]. Moreover, a detailed Raman study [29] has shown that the low frequency vibration of R-ions in $RB_6$ cannot be regarded as an independent mode, but as a coherent mode with long-range interaction between R-ions. Taking these findings into account we have developed here a model in which the role of independent oscillators is assigned to the Gd–Gd dimers. The dimer components, on the contrary, lose their independence, and their displacements may be anharmonic. To test this hypothesis, we re-refined the structure of $GdB_6$ at nine temperatures, taking into account the anharmonicity of Gd ions up to the sixth order (model I). The previous harmonic model was named II for comparison. The values of $u_{eq}(\text{Gd})$ from model I were assigned to $u_{eq}(\text{Gd–Gd})$ and substituted into Eq. (2) together with doubled atomic mass $2m_a(\text{Gd–Gd})$ to estimate the Einstein temperature. Similar estimation was done from model II using corresponding $u_{eq}(\text{Gd})$ and a single atomic mass $m_a(\text{Gd})$. The resulting $u_{eq}(\text{B})$ values were almost equal for both models. Those from model I were then substituted into (3) to estimate the Debye temperature.

The set of parameters deduced by the analysis based on Eqs. (2)-(3) in models I and II is presented in Table 1. It contains values of $\theta_E$ and $\theta_D$ as well as zero temperature atomic vibrations $<u^2>_{zero}$ obtained from Eqs. (2) and (3) at $T = 0$, static components $<u^2>_{shift}$, and summary temperature independent components $<u^2>_{sum} = <u^2>_{shift} + <u^2>_{zero}$. As can be seen from Table 1, the refined value of the Debye temperature $\theta_D$ (I) = 1206 ± 48 K coincides within the limits of accuracy with $\theta_D$ = 1160 K derived from heat capacity measurements (Fig. 2c). The Einstein temperature $\theta_E$ (I) ≈ 80 K determined on the assumption of Gd–Gd dimers is less than $\theta_E$ (II) ≈ 90 K but, in general, it corresponds to the energy of 7 ÷ 8 meV of the Gd quasilocal mode found in [26- 28]. For Gd ions in model I, the contribution of the static component to the temperature independent atomic displacement parameter $<u^2>_{sum}$ is an order of magnitude higher than that from $<u^2>_{zero}$, while two these contributions are almost equal for the B atoms. Moreover, in the case of vibrationally coupled Gd dimers (model I) the total (static plus dynamic) mean-square displacements of the heavy Gd atom ($m_a$ ≈ 157 amu) exceed significantly those of the light B atom ($m_a$ ≈ 11 amu) (Fig. 5a, b). The same situation was discussed in detail in [22] where the inequality $u_{eq}(B) < u_{eq}(R)$ was found to be valid for all studied RE hexaborides ($R$ = La, Ce, Pr, Nd, Sm, Gd, Eu).

**3.4. Electronic and lattice instability in GdB$_6$.** When discussing the nature of (*i*) the large amplitude atomic displacements in GdB$_6$, (*ii*) the formation of the dynamically coupled Gd–Gd dimers and (*iii*) the appearance of dynamic charge stripes at low temperatures observed in the present study for the first time, it is worth noting also the recent results of dynamic conductivity investigation of Gd$_x$La$_{1-x}$B$_6$ [16]. In particular, it has been found in [16] that there are two components in the dynamic conductivity spectra, and, additionally to the contribution from Drude electrons a strong collective mode has been observed with a frequency of ~ 1000 cm$^{-1}$ and with a damping of 2200 cm$^{-1}$ (overdamped oscillator), which includes up to 70% of the conduction electrons available in these metals. The collective mode is typical for systems with cooperative dynamic Jahn-Teller (JT) effect in boron clusters [33-34] and in the case of GdB$_6$ it results from JT instability of B$_6$ molecules. In this scenario the cooperative high-frequency JT boron vibrations cause the rattling modes of Gd ions, which are quasilocal low-frequency vibrations (Einstein mode). The Einstein oscillators are characterized by very large vibration amplitude that leads to strong variation of the 5*d*–2*p* hybridization of electronic states of Gd and boron ions. According to the results of band structure calculations for RE hexaborides [35-37] the Gd *5d* and B *2p* states contribute to the conduction bands. Thus, the modulation of the conduction band produces "hot charge carriers", which in turn are strongly scattered on the Gd quasilocal mode. With the temperature lowering these oscillating hot electrons in GdB$_6$ form a filamentary structure (dynamic charge stripes), which is detected in present study and shown on the ED maps of Fig. 4. As shown in Fig. 4, a structure of nanometer size AC-conducting channels is formed at low temperatures in the (100) plane of GdB$_6$. The channels are oriented approximately in the [001] direction and they are accompanied with formation of vibration-coupled Gd-pairs. As a result, the strongest charge carriers scattering in GdB$_6$ is due to (i) the boron sub-lattice JT instability and (ii) the appearance of the non-equilibrium (hot) electrons which involved in the high frequency (~240 GHz [34]) quantum motion in dynamic charge stripes.

**4. Conclusion.** Fine details of the crystal and electronic structure have been studied in combination with features of the atomic dynamics in $GdB_6$, which is located in vicinity of the quantum diffusion regime of the charge transport and demonstrates a wide range linear temperature dependence of resistivity. As a result of high-precision X-ray diffraction studies of high-quality single crystals of $GdB_6$ in the temperature range 85-300K, it has been shown that already at room temperature Gd ions combine into dynamically coupled Gd–Gd dimers. With the temperature decrease, the emergence of dynamic charge stripes is observed in the vicinity of the (100) planes forming a nanometer size filamentary structure of AC-conducting channels in the [001] direction. It has been concluded that just these two factors representing the lattice and electron instabilities are responsible for the extremely strong charge carrier scattering which leads to the diffusion regime of the charge transport in $GdB_6$. The size of Gd–Gd dimers is estimated to be ~3.3Å, and variation of the barrier height in the double-well potential is within the 3-8 meV limit.

**Acknowledgements.**

This work was supported by the Russian Ministry of Science and Higher Education within the State assignment FSRC 'Crystallography and Photonics' Russian academy of Science (RAS) in part of methodological developments for X-ray data processing and by the Russian Science Foundation, project no. 17-12-01426, in part of studying the structure and properties of $GdB_6$ crystals. X-ray diffraction data were collected on the equipment of the Shared Facility Center of FSRC 'Crystallography and Photonics' of RAS; measurements of heat capacity were carried out in the Center of Low Temperature Physics, Slovak Academy of Sciences. The work of K. Flachbart and S. Gabani was supported by the Slovak agencies VEGA (grant no. 2/0032/16) and APVV (grant no. 17-0020).

**Figure captions.**

**Fig.1.** (a) Unit cell of GdB$_6$. Difference Fourier maps of the residual electron density ($\Delta g$) distribution in the (100) plane of the crystal lattice at temperatures (b) 293 K, (c) 200 K and (d) 90 K.

**Fig.2.** Temperature dependences of (a) resistivity and (b) specific heat in GdB$_6$. For comparison, data obtained for the non-magnetic reference compound LaB$_6$ are also shown. The straight line approximates the linear part of the resistance curve in the range of 30 ÷ 300 K. Different contributions to heat capacity: (c) the Debye contribution $C_D$; (d) the Einstein contribution $C_E$; (e) the low temperature Schottky component $C_{Sch}$. Panels (f-g) show (f) a schematic view of the double-well potential with a barrier height $\Delta E$ and (g) the temperature dependence of barrier height derived from the heat capacity (triangle) and X-ray (circles) experiments.

**Fig. 3**. (a) Temperature dependences of interatomic distances Gd – B, (B – B)$_{intra}$ within B$_6$ octahedra and (B – B)$_{inter}$ between them in the cubic lattice of GdB$_6$. Temperature dependences of symmetry-independent lattice parameters *a*, *b*, *c* (b) and angles α, β, γ (c) of the GdB$_6$ crystal structure.

**Fig. 4.** Difference Fourier maps of residual ED ($\Delta g$) distribution at temperatures (a) 293K, (b) 200K and (c) 90K in three orthogonally related planes {100} of the crystal lattice passing through Gd atoms. Three blocks of eight unit cells each are shown for easier viewing of ED (electron density) filamentary structures.

**Fig.5.** Temperature dependences of mean-square atomic displacements for (a) boron $U_B$ and (b) Gd cation $U_{Gd}$ as obtained from models I and II (see text for details).

**Table 1**. Einstein $\theta_E$ and Debye $\theta_D$ temperatures, temperature independent components of the mean-square atomic displacements $\langle u^2 \rangle_{zero}$, $\langle u^2 \rangle_{shift}$, $\langle u^2 \rangle_{sum} = \langle u^2 \rangle_{zero} + \langle u^2 \rangle_{shift}$, refined in structure models I (anharmonic for Gd and harmonic for boron) and II (harmonic both for Gd and B); $R$ – approximation accuracy.

| Model (atoms) | $\theta_E / \theta_D$ K | $\langle u^2 \rangle_{zero}$, Å$^2$ | $\langle u^2 \rangle_{shift}$, Å$^2$ | $\langle u^2 \rangle_{sum}$, Å$^2$ | $R$, % |
|---|---|---|---|---|---|
| I (B) | - / 1206±48 | 0.0028 | 0.0020 | 0.0048 | 1.38 |
| I (Gd-Gd) | 79±3 / - | 0.00098 | 0.00906 | 0.01004 | 1.62 |
| II (Gd) | 90±1 / - | 0.00171 | 0.00050 | 0.00221 | 1.02 |

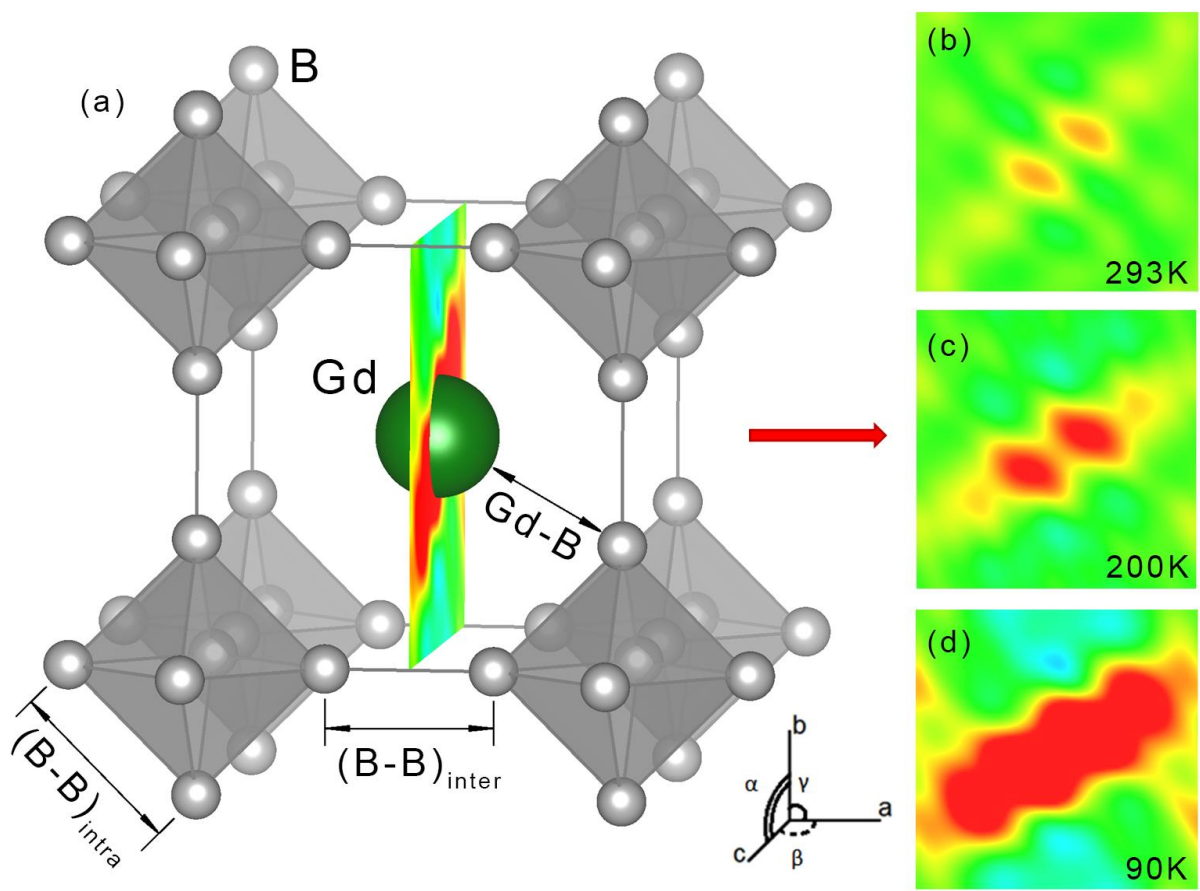

Fig. 1.

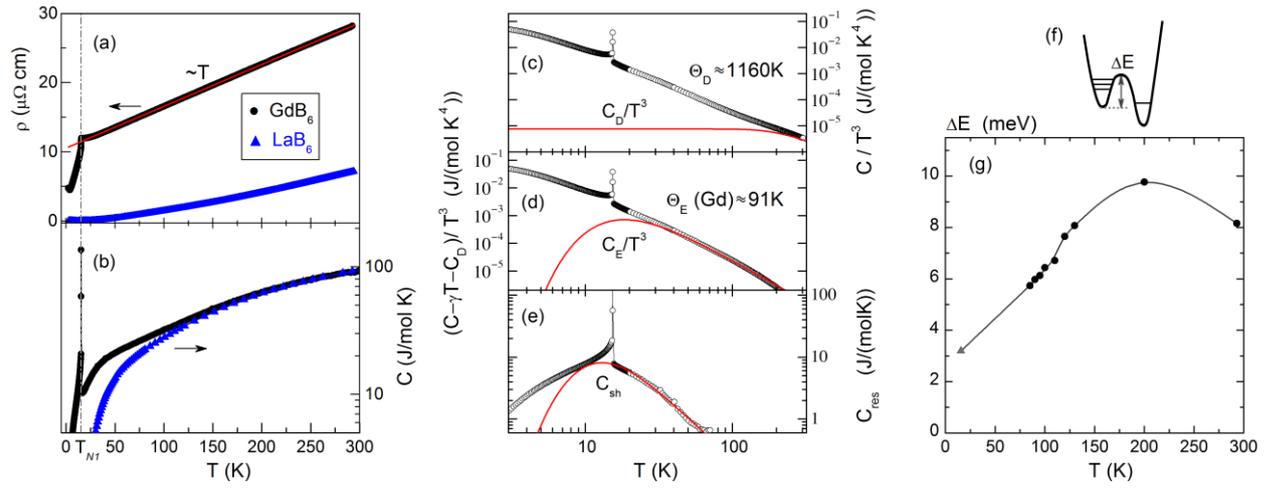

Fig.2.

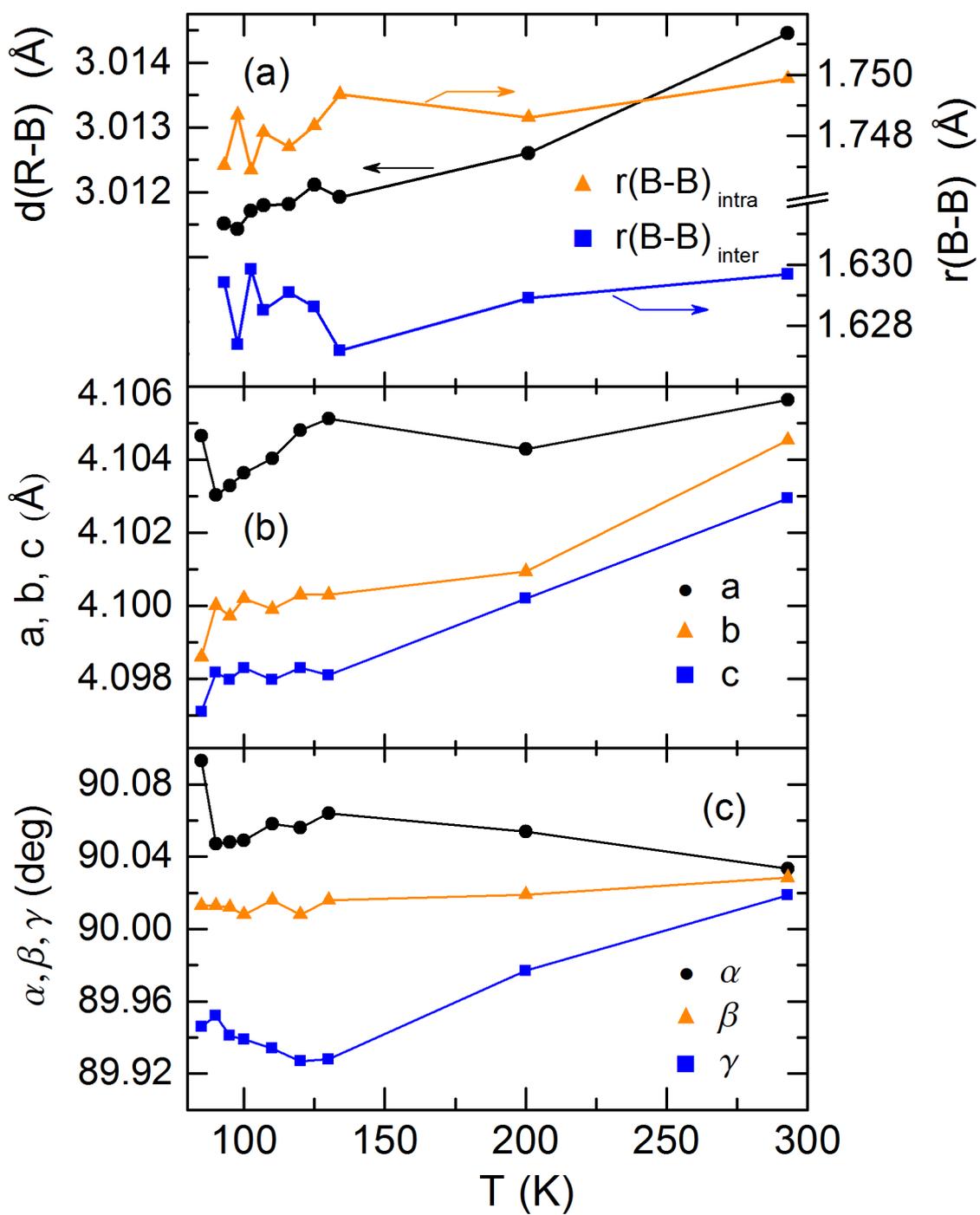

Fig. 3.

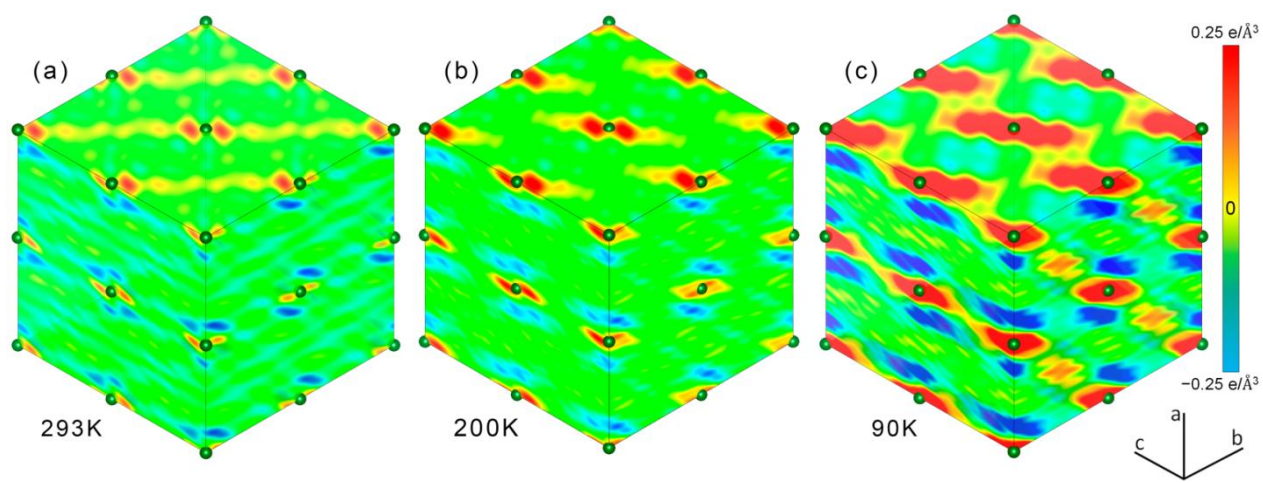

Fig. 4.

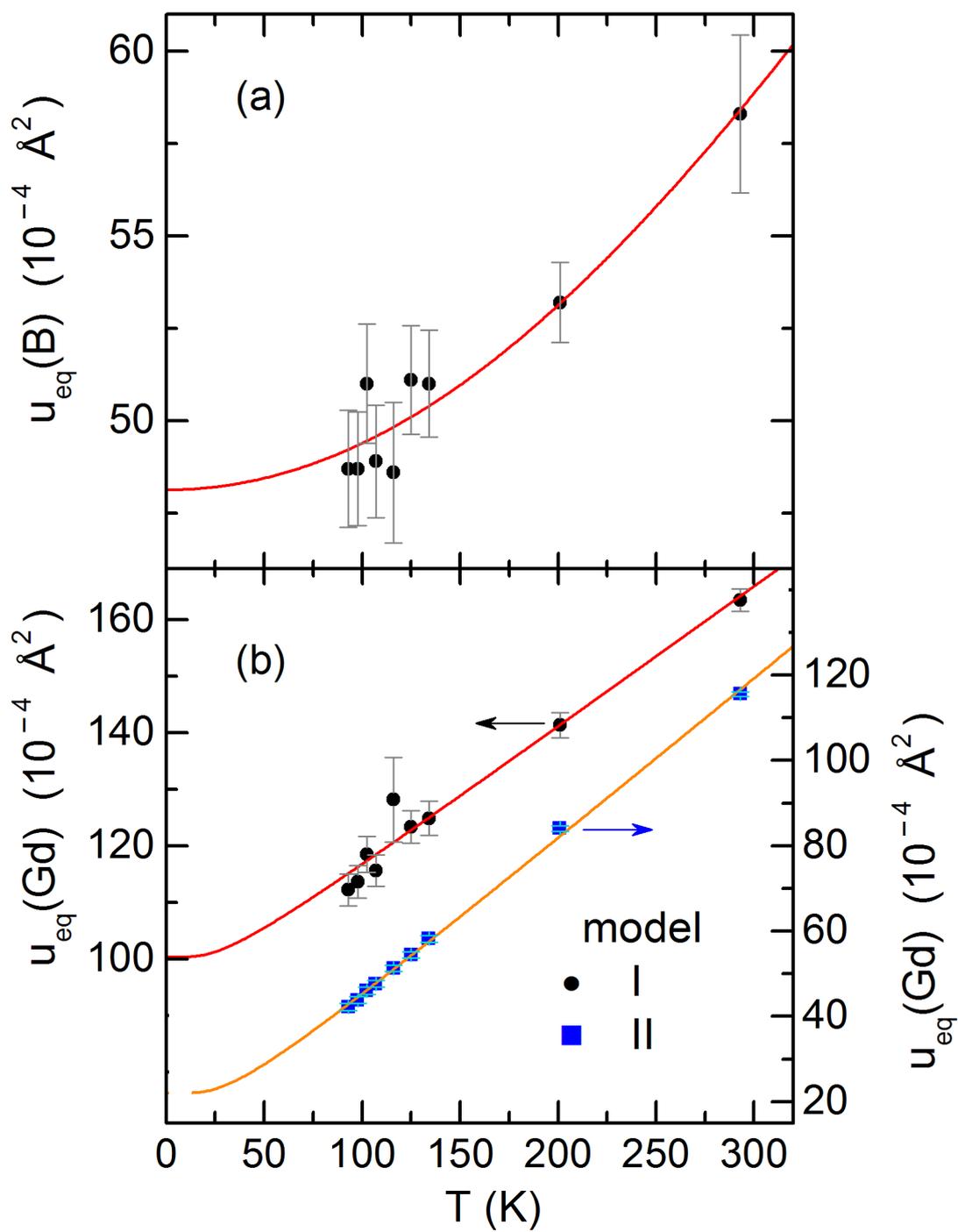

Fig.5.